\documentstyle[twoside,fleqn,espcrc2,epsfig]{article}

\title{Zero Temperature String Breaking with Staggered Quarks
\thanks{Presented by C.~DeTar.}}
\author{ C.~Bernard
\address{Department of Physics, Washington University, St.~Louis, MO 63130, USA},
T.~Burch
\address{Department of Physics, University of Arizona, Tucson, AZ 85721, USA},
T.A.~DeGrand
\address{Physics Department, University of Colorado, Boulder, CO 80309, USA},
C.E.~DeTar
\address{Physics Department, University of Utah, Salt Lake City, UT
  84112, USA},
Steven~Gottlieb
\address{Department of Physics, Indiana University, Bloomington, IN 47405, USA},
U.M.~Heller
\address{CSIT, Florida State University, Tallahassee, FL 32306-4120, USA},
P.~Lacock$\,\null^{\rm d}$,
K.~Orginos$\,\null^{\rm b}$,
R.L.~Sugar
\address{Department of Physics, University of California, Santa Barbara, CA 93106, USA},
and D.~Toussaint$\,\null^{\rm b}$,
} 
\begin{document}

\begin{abstract}
The separation of a heavy quark and antiquark pair leads to the
formation of a tube of flux, or ``string'', which should break in the
presence of light quark-antiquark pairs.  This expected
zero-temperature phenomenon has proven elusive in simulations of
lattice QCD.  In an extension of work reported last year we present
clear evidence for string breaking in QCD with two flavors of
dynamical staggered sea quarks and apply our results to a simple
three-state mixing model for string breaking.  We find that mixing is
weak and falls to zero at level crossing.
\end{abstract}

\maketitle 
\section{INTRODUCTION} 
The heavy quark-antiquark potential is known quite accurately in
quenched simulations\cite{ref:quenchedpot}.  It is traditionally
measured with the Wilson-loop observable, proportional to
$\exp[-V(R)T]$ at large $T$.  In the presence of dynamical quarks it
is expected that the potential levels off with increasing $R$,
signaling string breaking.  However, in QCD simulations string
breaking has proven to be very difficult to detect with this
observable, even out to $R \approx 2$ fm\cite{ref:fullpot1,ref:fullpot2}.

The reason string breaking has not been seen using the traditional
Wilson-loop observable has been apparent for some time
\cite{ref:michael,ref:DKKL,ref:drummond}. The Wilson loop can be
regarded as a hadron correlator with a source and sink state ``F''
consisting of a fixed heavy quark-antiquark pair and an associated
flux tube.  The correct lowest energy contribution to the Wilson-loop
correlator at large $R$ should be a state ``M'' consisting of two
isolated heavy-light mesons.  However, such a state with an extra
light dynamical quark pair has poor overlap with the flux-tube state,
so it is presumably revealed only after evolution to very large $T$.
To hasten the emergence of the true ground state, it is necessary to
enlarge the space of sources to include both F and at least one M
state.

String breaking has been demonstrated in the strong-coupling
approximation to QCD \cite{ref:drummond,ref:drummond2}, and a variety
of QCD-like theories
\cite{ref:Knechtli_Sommer,ref:Stephenson,ref:deForcrand_Philipsen,ref:Philipsen_Wittig,ref:Trottier,ref:Stewart_Koniuk},
and in SU(3) with dynamical quarks at nonzero
temperature\cite{ref:DKKL}.  Last year, we reported a preliminary
low-statistics result for two flavors of staggered quarks 
\cite{ref:lacock_lat99}, and, earlier this year, Pennanen and Michael
announced evidence for string breaking at zero temperature using
Wilson-clover quarks and a novel technique for variance reduction in
computing the light quark propagator \cite{ref:Pennanen_Michael}.
This year we report results with higher statistics \cite{ref:article}.

\section{METHODOLOGY}
We work with 198 configurations of size $20^3 \times 24$, generated
with the one-plaquette gauge action at $6/g^2 = 5.415$ in the presence
of two flavors of conventional dynamical staggered quarks of mass $am
= 0.0125$.  The lattice spacing ({\it via} the Sommer parameter
\cite{ref:SOMMER}) is approximately 0.163 fm, and $m_\pi/m_\rho =
0.358$, giving a comfortable box size and a relatively light quark.

\begin{figure}
\epsfig{bbllx=50,bblly=248,bburx=566,bbury=499,clip=,
 file=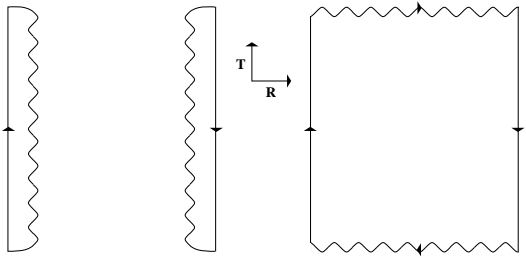,width=70mm}
\vspace*{-5mm}
\caption{The static-light meson-antimeson pair contribution
to the full QCD propagator. The wiggly lines denote
the light quark propagator. Shown are the `direct' and 
`exchange' terms respectively.
\label{fig:D1}
}
\vspace*{-5mm}
\end{figure}
\begin{figure}
\epsfig{bbllx=48,bblly=247,bburx=566,bbury=501,clip=,
 file=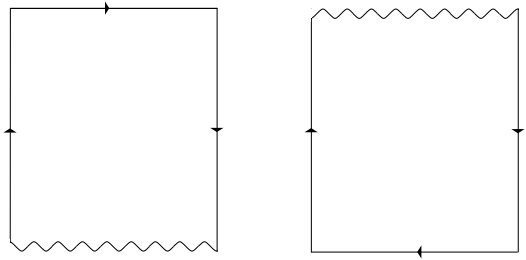,width=70mm}
\vspace*{-5mm}
\caption{The string-meson correlation matrix element $G_{FM}$
(and its hermitian conjugate $G_{MF}$).
The wiggly line again denotes the light quark propagator.
\label{fig:E1}
}
\vspace*{-7mm}
\end{figure}
Our flux tube ``F'' source and sink operator is constructed from the
product of smeared spatial links, for the most part taken along one of
the lattice axes, using 10 APE smearing iterations
\cite{ref:APEblock}, combining the direct link with a factor
$1-\alpha$ (in our case, $\alpha = 0.294$) and six staples with factor
$\alpha/6$ with SU(3) projection after each iteration.  The correlator
of this operator $G_{FF}(R,T)$ is the familiar Wilson loop with
smeared space-like segments, and point-like static quark and antiquark
lines.  Our two-meson source and sink operator ``M'' is the direct
product of static-light meson and antimeson operators.\footnote[1]{In
the continuum limit the meson pairs created by this operator are in a
combination of light quark singlet and nonsinglet states, and only the
singlet part should mix with the flux tube.  On our coarse lattice,
where only one meson-meson state is important, this affects the
overlap $Z_M$, but not the mixing parameters.  In this analysis we
have not attempted to reconcile the restriction to two flavors in
internal quark loops with the flavor counting of the external states,
which would affect the relative weights of the ``direct'' and
``exchange'' contributions to $G_{MM}$.  With the accuracy we have
achieved so far, this makes no detectable difference.}  In
constructing the static-light meson we use a light-quark wave function
with weight 2 at zero separation and 1 on each of the six on-axis
second neighbors.  Thus we measure the additional correlation matrix
elements $G_{MM}(R,T)$, $G_{MF}(R,T)$ and $G_{FM}(R,T)$, diagramed in
Figs.~\ref{fig:D1} and \ref{fig:E1}.  These observables are computed
for all translational and cubic rotational displacements.  The
required all-to-all propagator is estimated using a random source
technique with 128 random sources per configuration.

\begin{figure}
 \vspace*{-5mm}
 \epsfig{bbllx=0,bblly=0,bburx=4000,bbury=4000,clip=,
 file=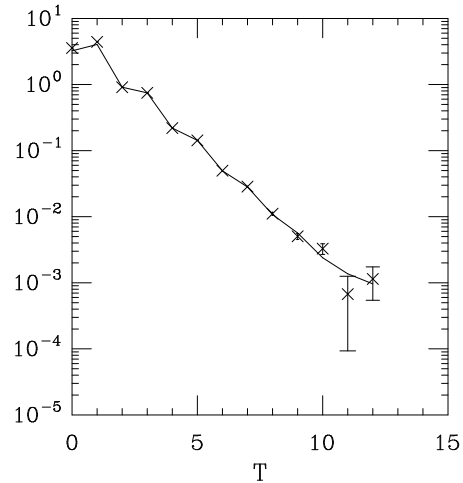,width=70mm}  \vspace*{-10mm}
\caption{Static-light propagator with a nonoscillating $S$-wave and
oscillating $P$-wave component. The solid line connects the best fit
values.  \label{fig:slprop}
}
 \vspace*{-8mm}
\end{figure}

\section{RESULTS}
The static-light meson correlator in Fig.~\ref{fig:slprop} has
contributions from both a nonoscillating $S$-wave ($B$-meson-like)
light-quark orbital and oscillating $P$-wave ($B^*$-meson-like)
orbital with $aE_S = 0.7884(12)$ and $aE_P = 1.022(6)$.  Thus our
two-meson correlator includes nonoscillating combinations $SS$ and
$PP$ and an oscillating $SP$.  If mixing is weak, our two-channel
correlators should also include a flux-tube level with an
approximately Coulomb-plus-linear behavior in $R$.  Our fit ansatz for
the two-channel correlator is
\begin{equation}
   G_{AB}(R,T) = \sum_{i=1}^N Z_{Ai}^*(R) Z_{Bi}(R)[\lambda_i(R)]^{T+1} ~,
\label{eq:transition}
\end{equation}
where $A,B$ refer to the flux tube $F$ or meson-meson $M$ states and
$\lambda_i(R)$ is a real (positive or negative) eigenvalue of the
transfer matrix.  The channel energy is $-\log(|\lambda_i(R)|)$.

For simplicity we restrict the analysis to three spectral components
\begin{eqnarray}
\lambda_1(R) &=& e^{-V_1(R)} \nonumber \\
\lambda_2(R) &=& -e^{-V_2(R)}  \label{eq:fit_ansatz} \\
\lambda_3(R) &=& (-)^{R+1}e^{-V_3(R)} ~, \nonumber 
\end{eqnarray}
corresponding in the unmixed language, respectively, to the $SS$,
$SP$, and flux-tube components.  The peculiar phase for the flux tube
component is required by discrete lattice symmetry and can be obtained
simply by interpreting the Wilson loop as a heavy staggered quark
loop.

\begin{figure}
 \vspace*{-5mm}
 \epsfig{bbllx=0,bblly=0,bburx=4000,bbury=4000,clip=,
 file=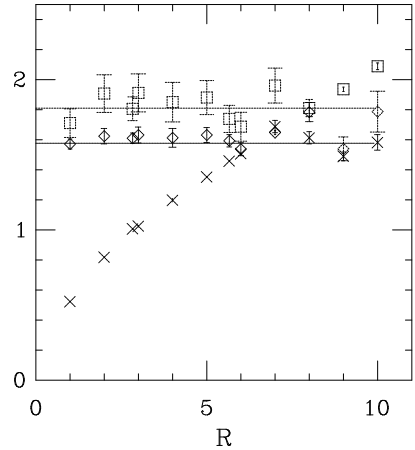,width=70mm} \vspace*{-5mm}
\caption{Heavy quark potential and first two excited states {\it vs}
separation $R$.  The dashed and solid lines give the asymptotic values
$2aE_S$ and $E_P + E_S$.  Jackknife errors are shown.
\label{fig:potential}
}
\vspace*{-5mm}
\end{figure}
Results for the three levels are plotted in Fig.~\ref{fig:potential}.
We see clear evidence for string breaking at the first level crossing
(about 1 fm), but with our statistics, we see no evidence for rounding
associated with avoided level crossing.  Thus string-breaking
pair-creation and annihilation transitions are weak in QCD.

\begin{figure}
 \vspace*{-5mm}
 \epsfig{bbllx=0,bblly=0,bburx=4000,bbury=4000,clip=,
 file=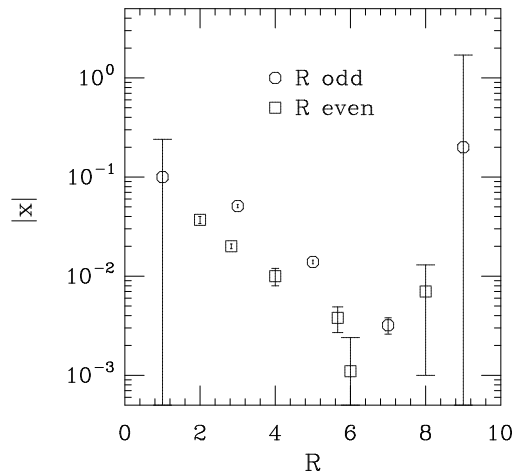,width=70mm}
 \vspace*{-5mm}
\caption{Absolute value of the mixing parameter $x$ {\it vs}
separation $R$.  Odd and even series are distinguished.
\label{fig:mixing_x} 
}
\vspace*{-5mm}
\end{figure}
\begin{figure}
 \vspace*{-6mm}
 \epsfig{bbllx=0,bblly=0,bburx=4000,bbury=4000,clip=,
 file=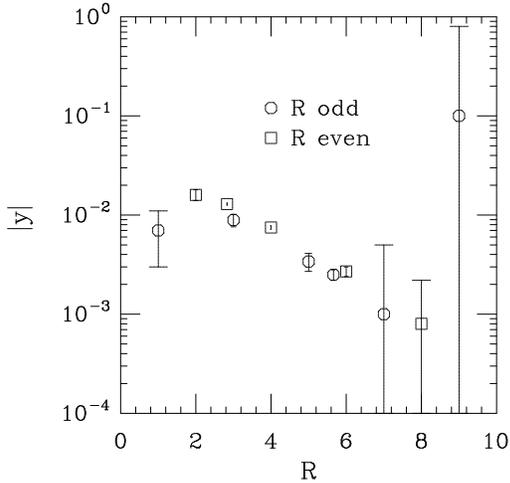,width=70mm}
 \vspace*{-12mm}
\caption{Absolute value of the mixing parameter $y$ {\it vs}
separation $R$.  Odd and even series are distinguished.
\label{fig:mixing_y} 
}
\vspace*{-8mm}
\end{figure}

To explore mixing further, we introduce a mixing model similar to that
of Drummond, Pennanen and
Michael\cite{ref:drummond,ref:Pennanen_Michael}.
\begin{equation}
   G(R,T) = {\tilde Z}^0(R) {\cal T}(R)^{T+1} Z^0(R) ~,
\end{equation}
where the transfer matrix is given by
\begin{equation}
    {\cal T}(R) = \left(\begin{array}{ccc}
        \lambda^0_1(R) & 0 & x \\
        0 & \lambda^0_2(R) & y \\
        x & y & \lambda^0_3(R)  
    \end{array} \right)
\end{equation}
and the mixing coefficients are $x$ and $y$.  This model matches our
fit ansatz reasonably well, allowing us to extract values for the
mixing parameters, the {\em magnitudes} of which are plotted in
Figs.~\ref{fig:mixing_x} and \ref{fig:mixing_y}.  We see that mixing
is weak and is small in the vicinity of level crossing.

\section{DISCUSSION}

To see string breaking in the heavy-quark potential, it is helpful to
introduce explicit meson-meson channels.  Mixing among the channels
is found to be very weak.  String breaking occurs at the first level
crossing, {\it i.e.}, with our quark mass, about one fermi.  These results
justify a two-channel model of excited quarkonium decay, in which a
closed channel populated by bound states couples weakly to an open
channel \cite{ref:Drummond_Horgan}.

This work is supported by the US National Science Foundation and
Departmet of Energy and used computer resources at Indiana University,
the San Diego Supercomputer Center (NPACI), and the University of Utah
(CHPC).

\end{document}